# Properties of comet 9P/Tempel 1 dust immediately following excavation by Deep Impact

**Lev Nagdimunov[1], Ludmilla Kolokolova[1*], Michael Wolff[2], Michael F. A'Hearn[1], Tony L. Farnham[1]**

1. University of Maryland, Department of Astronomy, MD, USA
2. Space Science Institute, Boulder, CO, USA

* Corresponding author. tel. +1(301)405 -1539, fax: +1(301)405-3538,
Email: ludmilla@astro.umd.edu (Ludmilla Kolokolova)

## Abstract

We analyzed Deep Impact High Resolution Instrument (HRI) images acquired within the first seconds after collision of the Deep Impact impactor with the nucleus of comet 9P/Tempel 1. These images reveal an optically thick ejecta plume that casts a shadow on the surface of the nucleus. Using the 3D radiative transfer code HYPERION we simulated light scattering by the ejecta plume, taking into account multiple scattering of light from the ejecta, the surrounding nuclear surface and the actual observational geometry (including an updated plume orientation geometry that accounts for the latest 9P/Tempel 1 shape model). Our primary dust model parameters were the number density of particles, their size distribution and composition. We defined the composition through the density of an individual particle and the ratio of its material constituents, which we considered to be refractories, ice and voids. The results of our modeling indicate a dust/ice mass ratio for the ejecta particles of at least 1. To further constrain the parameters of the model, we checked for consistency between the ejecta mass resulting from our modeling with the ejecta mass estimated by the crater formation modeling. Constraining the particle size distribution by results of other studies of the Deep Impact ejecta, we find the number density of ejecta particles equal to ~$10^4$ particles/$cm^3$ at the base of the plume.

## 1. Introduction

In July 2005, the Deep Impact spacecraft reached comet 9P/Tempel 1 and released a 366 kg impactor that collided with the cometary nucleus at a speed of 10.3 km/s. The spacecraft observed the impact with its High Resolution (HRI) and Medium Resolution (MRI) Instruments. The observational coverage of the impact was obtained in two discrete windows: the first lasting for approximately 800 seconds after the collision; and the second as a look-back phase beginning approximately 45 minutes after collision. These images show that the collision produced an optically thick ejecta cloud composed of material excavated from the cometary nucleus. In fact, cloud opacity was sufficient to completely obscure the impact crater, as well as to cast a shadow on the nucleus.

For the analysis performed in this paper we used the HRI data taken from the first seconds after the impact (McLaughlin et al., 2006) in a clear filter (Klaasen et al., 2013) that centers at 650 nm (this wavelength was used in the following modeling). These data allow us to study the early stages of the ejecta cloud development, when one would expect to observe pristine material from the cometary nucleus (dust before it gets processed, sublimated or fragmented). Additionally, the use of a narrow temporal range simplifies the analysis by restricting the variations in the viewing geometry. Examples of the HRI images used for our analysis are shown in Fig. 1. One can clearly see the shadow above the bright ejecta plume. The complex structure of the shadow (see Kolokolova et al., 2012) provides the opportunity to study changes in the cloud opacity with height and azimuth, revealing changes in the characteristics of the excavated material as the impact crater is developing, i.e. depending on the depth and surface features from which the ejecta originated. Furthermore, the use of image comparison allows one to characterize the temporal evolution of the shadow and the ejecta cloud, thus providing direct insight into the nuclear structure (changes in the “bottom” of the shadow) and evolution of the ejecta (changes in the “top” of the shadow).

In this paper we introduce the method that we are planning to use in the future for a detailed study of shadow structure and evolution and provide some tentative results that characterize the ejecta cloud at the first second after the impact, when the ejecta composition is closest to undisturbed comet material. We thereby demonstrate the utility of shadow analysis for characterizing properties of the cometary nucleus. Currently, putting aside the inhomogeneity of the shadow, we check if shadow characteristics allow us to reveal properties of the excavated material and analyze what characteristics of the ejecta (e.g. composition, size distribution, number density) can be revealed through analysis of the shadow. The obtained information can then be applied to a detailed study of the shadow's variation. However, in this paper we limit ourselves to analysis of primarily one image taken 1.036 seconds after Deep Impact (image (a) in Fig. 1).

## 2. Dynamical and light-scattering modeling of the ejecta cloud

A thorough analysis of the ejecta cloud characteristics requires one to explicitly treat the three dimensional and multiple scattering nature of the impact event: the geometry of the ejecta cloud; the geometry of its illumination and observation; the reflection from the nucleus; and the extinction and scattering of solar radiation (both direct and reflected) by the cloud. To accomplish this, we employed the recently released 3-D Monte Carlo radiative transfer package named HYPERION (Robitaille, 2011). In addition to the fundamental radiative transfer algorithms, this package provides a Python-based flexible pre-processing system that allows one to implement an irregularly-sampled grid in multiple coordinate systems and to employ arbitrary opacity distributions with multiple dust components. These features were used to represent explicitly the non-spherical shape of the cometary nucleus as well as the inhomogeneous nature of the ejecta plume. HYPERION also includes the capability for parallelization, which we exploited using the NASA-Ames "Pleiades" cluster.

The shape of comet Tempel 1 nucleus was ultimately specified using the shape model from Farnham and Thomas (2013). Our initial attempts to use a spherical nucleus revealed that we could not reproduce the shadow with a simpler model. The “flatter” nature of the nucleus at the impact site was not compatible with the use of a sphere. In essence, the curvature of the sphere “hid” the majority of the shadow. Although our simulations were run on the Pleiades cluster, we found that it was necessary to use an irregularly spaced grid in order to avoid exceeding the available RAM, i.e., specifying a uniform grid that adequately resolved the plume was prohibitive. However, even in our final prescription, we required approximately 30 GB of RAM per processing. The ultimate grid, in high precision areas (around and within ejecta cone and shadow), was defined in spherical coordinates with individual grid cell sizes of 6.7 meters in the radial direction, 0.6º in the azimuthal direction and 0.7º in the polar direction.

In order to specify the plume properties, we needed to define geometric characteristics of the ejecta plume together with illumination and observation angles. Starting with the latter, we specified the positions of the Sun and spacecraft with respect to the Tempel 1 coordinate system defined in Farnham and Thomas (2013). Deep Impact SPICE data (Semenov and Acton, 2006) indicate that the sub-spacecraft point had a longitude 275.66° and latitude -40.97°, while the sub-solar longitude was 332.76° and latitude was -4.02°. The phase angle for our observations was 62.93°. To determine an appropriate plume shape, we considered the dynamical modeling of the impact and crater development by Richardson et al. (2007). They determined that the impactor had to impact the nucleus at an oblique angle (with respect to the surface) in order to produce the shape and spread of the ejecta plume and its shadow as observed by the Deep Impact instruments. Following their recommendation, we represented the ejecta plume as a cone centered at the impact point on the nucleus. To reduce the plume orientation parameter space that we needed to explore in our initial investigation, we adopted the Richardson et al. orientation as a starting point. They defined their plume orientation by specifying the longitude and latitude of the impact site (apex of the cone, at 15° and -26°) and the North-South and East-West tilt angles of the cone's axis, relative to the surface normal at that point (0° North, 45° West). However, between the time of their study and our work, the Tempel 1 nucleus shape model was updated (Thomas et al., 2013), along with its corresponding axis orientation and surface coordinates. Using the cometocentric reference frame for the new model thus changed the location of the impact site as well as the specified orientation of the cone. Because we used the improved shape, we needed to convert the Richardson et al. orientation into the new model coordinates. To do this, we used the DI SPICE information describing the viewing geometry, along with the original shape model, to convert the Richardson et al. cone orientation into the J2000 reference frame. Because this is an inertial frame, it is independent of any particular shape model, which allows it to be used as an intermediate step for converting the orientation into the coordinate system of any shape model that is desired. For the new Tempel 1 shape model, the impact site longitude and latitude are 16° and -28°, respectively, and the tilt angles of the Richardson et al. cone are 58.2° West and 2.3° North relative to the surface normal at the impact point. However, our model required the orientation with respect to the cometocentric normal at the impact site (e.g., for purposes of generating the cone particles, the surface of the shape model was ignored and the orientation was described relative to the radial vector). Converting the J2000 coordinates into this frame gives an azimuthal angle of 317° and a tilt angle of 62°, which acted as our starting point for the modeling. After iterating through a series of simulation-observation comparisons, this cone orientation was adjusted until we found the "best fit" parameters of 310° and 75° respectively (see Fig. 2). Thus, the orientation of the cone in our final model differs from that in the Richardson et al. model by ~13.2°.

In terms of light scattering properties required for the radiative transfer calculations, we represented the nucleus as a surface formed by very densely located scatterers with a single scattering albedo of approximately 0.065 and whose phase function can be represented as a Henyey-Greenstein function with asymmetry parameter of -0.48 that we took from (Li et al., 2007, 2013). We found that for our simulations a single scattering albedo of 0.065 better reproduced the measured brightness at multiple points on the nucleus than the value 0.043 estimated by Li et al. (2007, 2013) based on Hapke modeling. This should not be surprising as classic radiative transfer is known to not apply well to such densely packed media as surfaces, and therefore our result is not presented here as an accurate albedo of the nucleus but rather an albedo that in our radiative transfer modeling reproduces its photometric properties as they are seen in the HRI images.

The primary adjustable parameters of our radiative transfer analysis of the plume are the single scattering albedo (SSA) and the extinction coefficient, $\alpha = n * C_{\text{ext}} = \rho * \chi_{\text{ext}}$ , where *n* is the number density of ejecta particles, $C_{\text{ext}}$ is the size-averaged extinction cross-section, ρ is the mass density of the ejecta and $\chi_{\text{ext}}$ is the mass extinction coefficient. As is well known, the extinction coefficient, $\alpha$, is related to the optical depth via $\tau = \int \alpha * ds$, where s is the path length along the beam. Since comet observers prefer working in units of number density and extinction cross-sections, while HYPERION typically accepts input as a function of mass, we can solve for $\chi_{\text{ext}} = \frac{n * C_{\text{ext}}}{\rho} = \frac{C_{\text{ext}}}{m_{\text{part}}}$ , where the mass density, $\rho = n * m_{\text{part}}$ , is the number density multiplied by the mass per individual particle. HYPERION requires the constituents of $\alpha$ separately, therefore we calculated and input into HYPERION $\chi_{ext}$ and ρ as two separate parameters computed via the equations above.

Thus, our fundamental plume radiative input parameters were the single scattering albedo (SSA), asymmetry parameter ($g$), number density ($n$), extinction cross section ($C_{\rm ext}$) and mass per individual particle ($m_{\rm part}$). To calculate SSA, $g$, and $C_{\rm ext}$ we considered the ejecta dust to be an ensemble of spherical (Mie) particles that obey a power law size distribution, i.e. number of particles within the range of sizes $r$ to $r+dr$ is proportional to $r^{-a}dr$. We then computed the size-averaged asymmetry parameter and the extinction and scattering cross sections for the size distributions presented in Table 1 and defined through the exponent of the power-law size distribution and the distribution cutoffs (minimum and maximum size of particles) using Eq. 1,

$$\mathrm{A} = \frac{\int_{r_{\min}}^{r_{\max}} \mathrm{A}(r) * r^{-a} dr}{\int_{r_{\min}}^{r_{\max}} r^{-a} dr} \qquad (1)$$

Where A can be $g$, $C_{\rm ext}$ and $C_{\rm scat}$. The size distribution particle size range and power law values were selected from previous studies of the Deep Impact ejecta (see, e.g., Lisse et al., 2005, 2006; Sunshine et al., 2007; Bonev et al., 2009). SSA was calculated as the ratio of the size-averaged scattering cross section to the size-averaged extinction cross section. The size-averaged mass was computed using Eq. 1 with A(r) equal to mass of particle, $m_{\rm part} = 4/3\ \pi * r^3 * \rho_{\rm p}$ where $\rho_{\rm p}$ is individual particle density of the material, defined as the mass of a single particle divided by its enclosing volume. We selected two values for $\rho_{\rm p}$: 0.4 g/cm$^3$ (the value of bulk density for the nucleus of comet Tempel 1; cf. Veverka et al., 2013) and 1.75 g/cm$^3$ used by Richardson et al. (2007). We consider these two values as representative of the extremes of ejecta material density (see discussion in Weissman et al. 2004). The individual particle density is a fundamental constraint in the determination of the complex refractive indices of the ejecta particles, which represent a basic input value for the Mie code. These refractive indexes were calculated via the Maxwell Garnet mixing rule (with the dominant material being the matrix and less abundant material being the inclusions) and the compositions described below. We considered an ejecta particle's composition as an intimate mixture of ice, voids, and some refractive material, where the latter was prescribed using so-called "Halley-like dust," i.e. dust with a composition consistent with findings of Giotto mission to comet Halley (Jessberger et al., 1988). For this dust, a particle includes silicates, amorphous carbon, and organics. For additional details see Mann et al. (2004), where the refractive index of this Halley-like material was calculated for wavelengths around 0.6 μm to be 1.98 + 0.48*i* and density of this material (excluding ice and voids) is shown to be 2.4 g/cm$^3$. Because of issues of non-uniqueness (the same density could correspond to different ratios of ice, voids and dust), we considered a volume fraction of ice in the dust/ice mixture from 0.1 to 1 and added as much void space (vacuum) as necessary to reproduce the desired individual particle density. The resulting refractive indexes, Mie-calculated radiative properties of the ejecta particles and $m_{\text{part}}$ estimations are summarized in Table 1. Each set of optical properties was evaluated against the observations and the "best fit" values were selected based on the closeness of the modeled and observed peak values of brightness of the shadow and ejecta and similarity of the general trends in changing brightness within the ejecta plume.

## 3. Results and discussion

As Table 1 indicates, the specific choice of individual particle density ($\rho_{\text{p}}$) has a very significant impact on the ability to constrain composition proportions. At low density, one finds relatively large changes in the single-scattering albedo, from 0.52 to 0.70. In contrast to this, **for** the high density case – because changes in the ice fraction must match similar (but opposite) changes in the void volume – the relatively unchanging average (effective) refractive index produces a smaller single scattering albedo range of 0.45 to 0.55. The size-averaged extinction cross-section does not vary significantly with composition but does depend strongly on the size distribution. From this, one may hope that determining which $C_{ext}$ fits the observed images best would allow one to constrain the average particle size. However, the extinction coefficient, and consequently the optical depth, are functions of the extinction cross-section multiplied by the number density. Thus, there can be a variety of combinations of $C_{\rm ext}$ and $n$ that produce the same fit for a given scattering phase function.

To provide an additional constraint for our modeling, we used the total mass of the ejecta, $\mathrm{M_{total}} = \rho * V_{\rm cone} = n * m_{\rm part} * V_{\rm cone}$ , where $n$ is the number density averaged over the cone height, *h;* we assumed a simple expansion law $n \sim n(0)*h^{-3}$, and $V_{\rm cone}$ is the volume of the cone representing the ejecta plume. We found that representing the ejecta plume by a hollow cone, as suggested by Richardson et al. (2007), did not reproduce the observations. Specifically, it produced a decrease of the brightness in the center of the plume that is clearly seen in the later HRI images but not at early stages of ejecta development (cf. Fig. 1). We interpret this as an indication that in the first second after the impact the ejecta filled most of the volume of the cone. Laboratory impact experiments (e.g. Strycker et al., 2013; Schultz et al., 2007, 2009; Hermalyn et al., 2013) provide evidence that at early stages of the crater development, the ejecta plume can be filled and the ejecta cone becomes hollow later. This can especially be expected for oblique impacts, where a hydrodynamic explosion occurring almost immediately after impact lifts all the material above the impactor and initially fills the typical hollow cone created by expanding shockwave. Calculating $V_{\rm cone}$, we employed an upper velocity range of 700-1000 meters per second for the ejecta material during the first few seconds in accordance with Richardson et al. estimates, and used this to determine the height of the plume at a given time.

Richardson et al. (2007) provided three scenarios for crater developments as a function of effective target strength (see Table 4 in Richardson et al.). Based on this characterization, they predicted three sets of values of crater diameter, time of crater development and total mass of ejecta. Recent estimations of the impact crater made using Stardust NExT images of comet Tempel 1 report a crater diameter of ~200m (Schultz et al., 2013). This diameter corresponds to the case of zero effective target strength in the Richardson et al. paper. Combining these two results, we can estimate that the total time of crater formation was T~600 sec, with the total excavated mass being equal to M ~ $3\times10^7$ kg. Note that there is a different estimation of the crater diameter that suggests it to be equal to 49 +/- 12 m (Richardson and Melosh, 2013). This is the case of effective target strength 1 kPa from Richardson et al. (2007) and the ejecta mass for this case is equal to ~$10^6$ kg. This value appears to be too low if we compare it with the data of Keller et al. (2007), who estimated that only water molecules, without other ejecta components, constituted (4.5-9)$\times10^6$ kg and the total mass of the ejecta could reach $10^7$-$10^8$ kg.

The scaling formula for gravity dominated cratering from the paper by Richardson et al. (2005) is $r = \left(\frac{\sqrt{2}}{2} * (1+\varepsilon) * C_e R^{\varepsilon} t\right)^{\frac{1}{1+\varepsilon}}$ , where *r* is the radius of the crater at time *t*, *R* is the final radius of the crater, and *ε* is a material constant ranging from 1.8 for rocks to 2.6 for sand. With this equation, one can see that the mass of the excavated material *m* at the time *t* and the total mass of the excavated material, M, are related as $\frac{m}{\mathrm{M}} = \left(\frac{t}{\mathrm{T}}\right)^{\frac{3}{1+\varepsilon}}$, where T is the total time of crater formation. Taking M=$3\times10^7$ kg and T=600 sec (see above), we can estimate the mass of the ejecta at one second after the impact as *m*= $3\times10^4 - 1.4\times10^5$ kg. This result is consistent with $2\times10^4 - 1\times10^5$ kg estimated by Holsapple and Housen (2007) for gravity dominated impact cratering of sand. Adopting an estimated mass of the ejecta one second after the impact, we can derive a range for $n*m_{part}$, which leaves us a more restricted selection of values of $C_{ext}$ for use in our simulations.

Figure 3 shows the images that correspond to the best-fit computations for the individual particle density ($\rho_p$) values of 0.4 and 1.75 g/cm$^3$. In both cases, we were able to match the peak brightness for both the ejecta and the shadow, whose values are 9 W/(m$^2$ sr µm) and 0.2 W/(m$^2$ sr µm), respectively. The best fit parameters for both images are listed in Table 2. For $\rho_p$ equal to 0.4 g/cm$^3$, the results show highly porous particles with approximately 10% dust and almost 20% ice by volume. For $\rho_p$ equal to 1.75 g/cm$^3$, we obtain low porosity particles with approximately 60% dust and between 38% and 8% ice by volume. Both results show a dust/ice mass ratio greater than 1, consistent with previous findings by Küppers et al. (2005) and Keller et al. (2007). It should also be noted that the number density and size distribution constraints appear to be the same in both cases. This results from the fact that the changes in extinction cross-section and the mass per particle are correlated (see Table 1): an increase in the extinction cross-section for a given size distribution produces an increase in the mass per particle and vice-versa. In other words, our results are not very sensitive to the particle size distribution. As such, we must prescribe the size information from other means, e.g. from estimates by Lisse et al. (2005, 2006) and Bonev et al. (2009). Specifically, the size distribution presented in Table 2 is in agreement with the conclusion about high contribution of one-micron sized particles in the near-infrared spectra of the Deep Impact ejecta (Sunshine et al., 2007) as this size corresponds to the average particle size of the selected size distribution. Note that for the density 1.75 g/cm$^3$ we had to use an asymmetry parameter g that is slightly smaller, a difference of 0.018-0.044, than the one produced by Mie calculations**.** However, this is within the uncertainty provided by the difference between ideal spherical particles in Mie theory and realistic nonspherical comet particles (cf. Fig. 10.20 and 10.21 in Mishchenko et al., 2002).

The results presented in this paper have demonstrated that one can fit the brightness of ejecta and the shadow by employing multiple-scattering radiative transfer and by adopting reasonable physical characteristics of the ejecta dust, which in all cases has a dust to ice mass ratio equal to or greater than 1. In addition, we also identified the model parameters that are most sensitive to the employed imaging data. Specifically, we found that the individual particle density of the ejecta material (which depends only on particle composition) and, as soon as the size distribution can be constrained by other studies, the ejecta particle number density are the strongest constraints of the model. As part of our work, we have updated the plume orientation geometry estimates from Richardson et al. (2007) for a newer Tempel 1 shape model created from combined observations of the Deep Impact and the Stardust spacecraft.

We intend to leverage the knowledge gained from this study in a future study that focuses on the inhomogeneity of the shadow and its associated temporal evolution. The goal of this continuing effort is to reveal additional compositional and structural inhomogeneity information about the cometary nucleus.

**Acknowledgment**. This work was supported by the NASA Planetary Mission Data Analysis Program, grant #NNX10AP31G. We are also grateful to Tom Robitaille for providing open access to his radiative transfer package HYPERION.

**Table 1. Compositional and optical characteristics of the ejecta material that were used to model ejecta cloud**

| $\rho_p$ (g/cm$^3$) | *Refractive Index* Re | Im | *Volume Fraction of Ice in Ice+Dust Mixture* | *Porosity* |
|---|---|---|---|---|
| **0.4** | 1.075 – 1.355 | 0.023 – 0.082 | 0.1 – 0.9 | 0.65 – 0.82 |
| **1.75** | 1.094 – 1.286 | 0.053 – 0.124 | 0.1 – 0.4 | 0.05 – 0.22 |

| *a* | $r_{min}$ (μm) | $r_{max}$ (μm) | SSA | $C_{ext}$ (μm$^2$) | *g* | $m_{part}$ (kg) |
|---|---|---|---|---|---|---|
| **$\rho_p$ - 0.4 g/cm$^3$** | | | | | | |
| 3.0 | 0.1 | 100 | 0.543 – 0.619 | 0.552 – 0.741 | 0.412 – 0.443 | $2.75\times10^{-15}$ |
| 2.5 | 0.1 | 100 | 0.523 – 0.554 | 4.87 – 5.18 | 0.479 – 0.505 | $4.56\times10^{-14}$ |
| 3.5 | 0.1 | 100 | 0.540 – 0.703 | 0.103 – 0.218 | 0.364 – 0.413 | $2.01\times10^{-16}$ |
| 3.0 | 0.1 | 1000 | 0.526 – 0.596 | 0.230 – 0.286 | 0.221 – 0.241 | $7.96\times10^{-15}$ |
| 3.0 | 0.01 | 1000 | 0.529 – 0.595 | 0.00024 – 0.00030 | 0.0019 – 0.0021 | $8.37\times10^{-18}$ |
| 3.0 | 0.01 | 100 | 0.539 – 0.616 | 0.0016 – 0.0022 | 0.0042 – 0.0046 | $7.96\times10^{-18}$ |
| **$\rho_p$ - 1.75 g/cm$^3$** | | | | | | |
| 3.0 | 0.1 | 100 | 0.490 – 0.515 | 0.586 – 0.714 | 0.418 – 0.444 | $1.2\times10^{-14}$ |
| 2.5 | 0.1 | 100 | 0.509 – 0.526 | 4.91 – 5.12 | 0.485 – 0.508 | $2.0\times10^{-13}$ |
| 3.5 | 0.1 | 100 | 0.449 – 0.492 | 0.125 – 0.206 | 0.370 – 0.398 | $8.8\times10^{-16}$ |
| 3.0 | 0.1 | 1000 | 0.497 – 0.511 | 0.240 – 0.281 | 0.224 – 0.240 | $3.48\times10^{-14}$ |
| 3.0 | 0.01 | 1000 | 0.493 – 0.505 | 0.00026 – 0.0030 | 0.0019 – 0.0021 | $3.66\times10^{-17}$ |
| 3.0 | 0.01 | 100 | 0.490 – 0.505 | 0.0017 – 0.0021 | 0.0042 – 0.0045 | $3.48\times10^{-17}$ |

**Table 2. Best fit characteristics of the ejecta that correspond to the images in Figure 3**

| $\rho_p$ - 0.4 g/cm$^3$ | $\rho_p$- 1.75 g/cm$^3$ |
|---|---|
| **Ejecta phase function asymmetry parameter**<br>g=0.44<br>**Ejecta composition**:<br>Volume fractions:<br>9-11% Halley dust, 16-20% ice, 74-72% voids<br>Dust/ice mass ratio: 1.0 - 1.6<br>**SSA**: 0.55-0.56<br>**Ejecta size distribution**<br>Power law exponent: 3<br>Smallest particle radius: 0.1 µm<br>Largest particle radius: 100 µm<br>**Ejecta number density at the bottom of the cone, *n(0)***<br>1.5×10$^4$ particles/ cm$^3$<br>**Ejecta mass (t~1s):**<br>1×10$^4$ kg | **Ejecta phase function asymmetry parameter**<br>g=0.40<br>**Ejecta composition**:<br>Volume fractions:<br>69-57% Halley dust, 38-8% ice, 23-5% voids<br>Dust/ice mass ratio: 3.6 - 21.6<br>**SSA**: 0.50-0.52<br>**Ejecta size distribution**<br>Power law exponent: 3<br>Smallest particle radius: 0.1 µm<br>Largest particle radius: 100 µm<br>**Ejecta number density at the bottom of the cone, *n(0)***<br>1.5×10$^4$ particles/cm$^3$<br>**Ejecta mass (t~1s):**<br>5×10$^4$ kg |

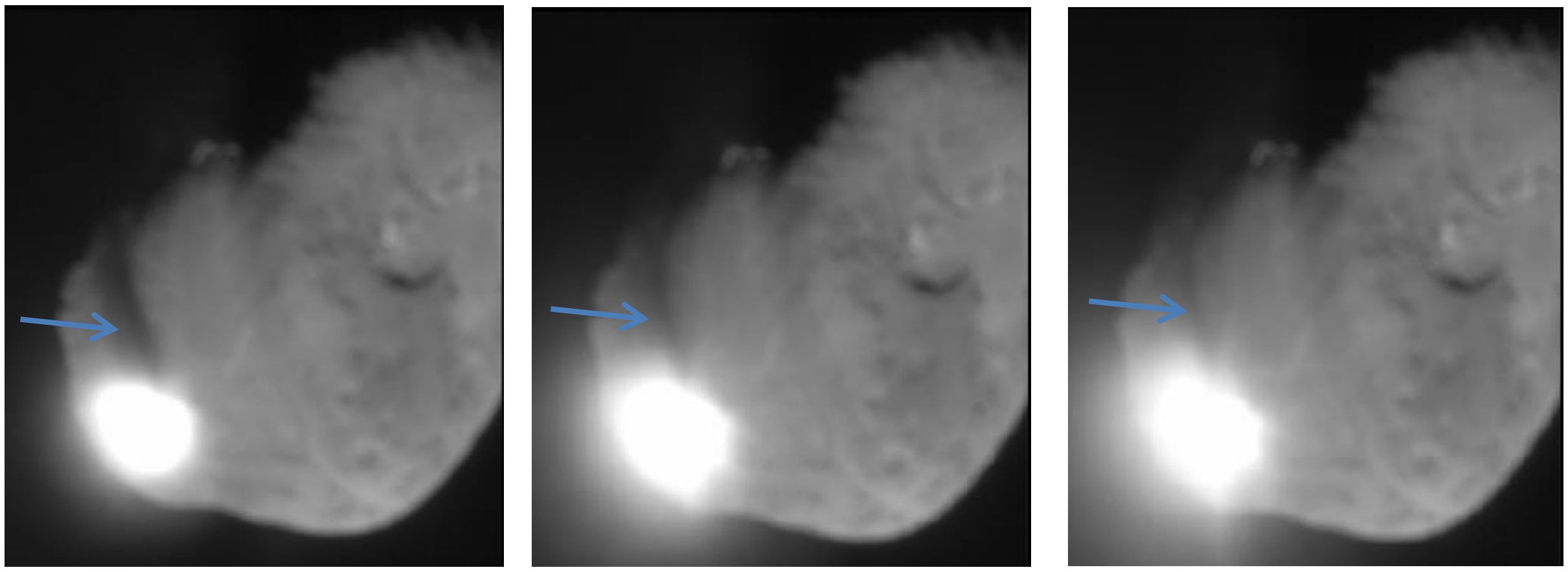

(a) (b) (c)

**Figure 1.** Development of the Deep Impact ejecta plume. The HRI images correspond to the following time after impact: (a) 1.036 s; (b) 1.880 s; (c) 2.723 s. An arrow in each image indicates the shadow cast by the ejecta plume.

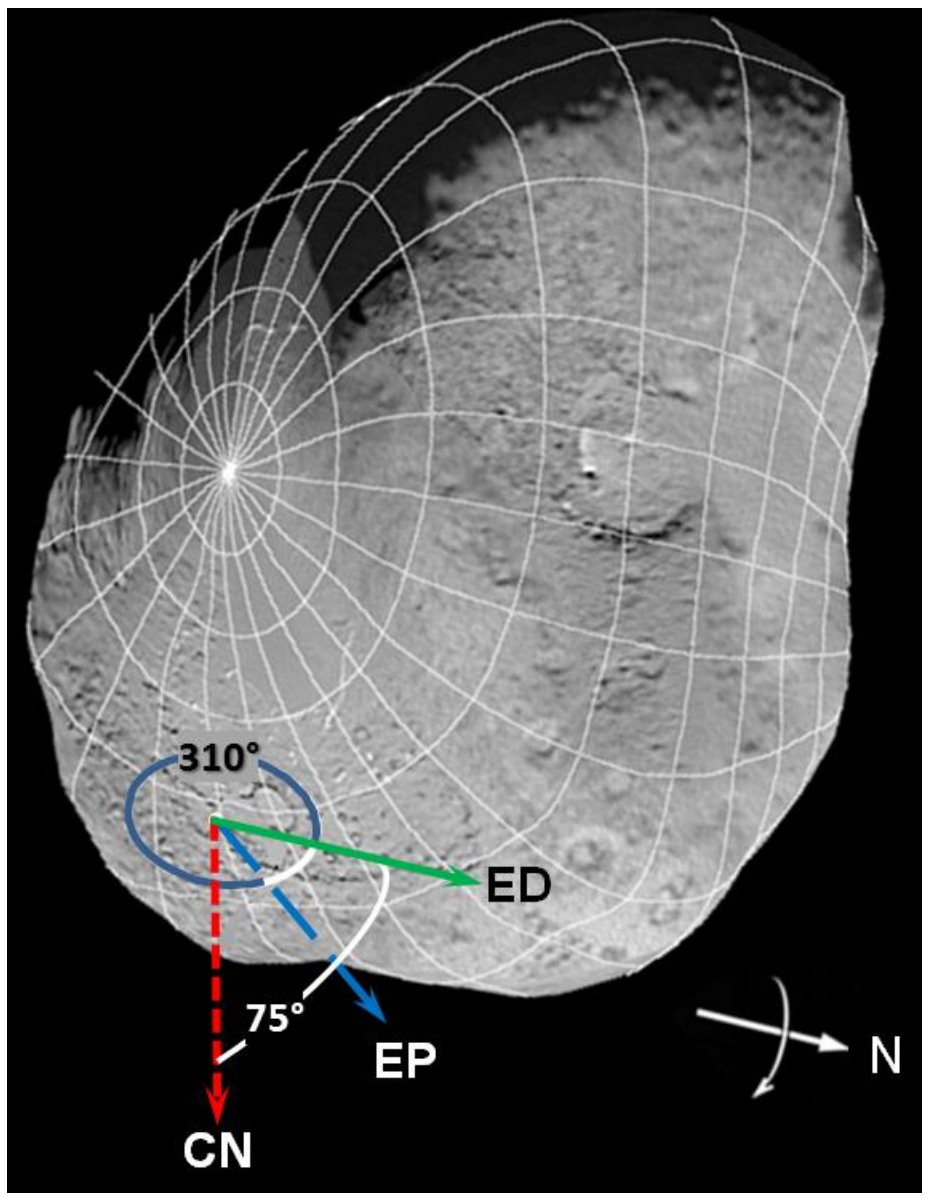

**Figure 2.** Geometry of Deep Impact ejecta used in our simulations. The ejecta plume direction (ED) is shown as a solid line; the projection of the ejecta (EP) onto the azimuth plane shown as a long-dashed line; cometocentric normal (CN) is shown as a short-dashed line. The ellipse is in the plane perpendicular to CN and its dark section represents the azimuth angle. Image adapted from Farnham and Thomas (2013).

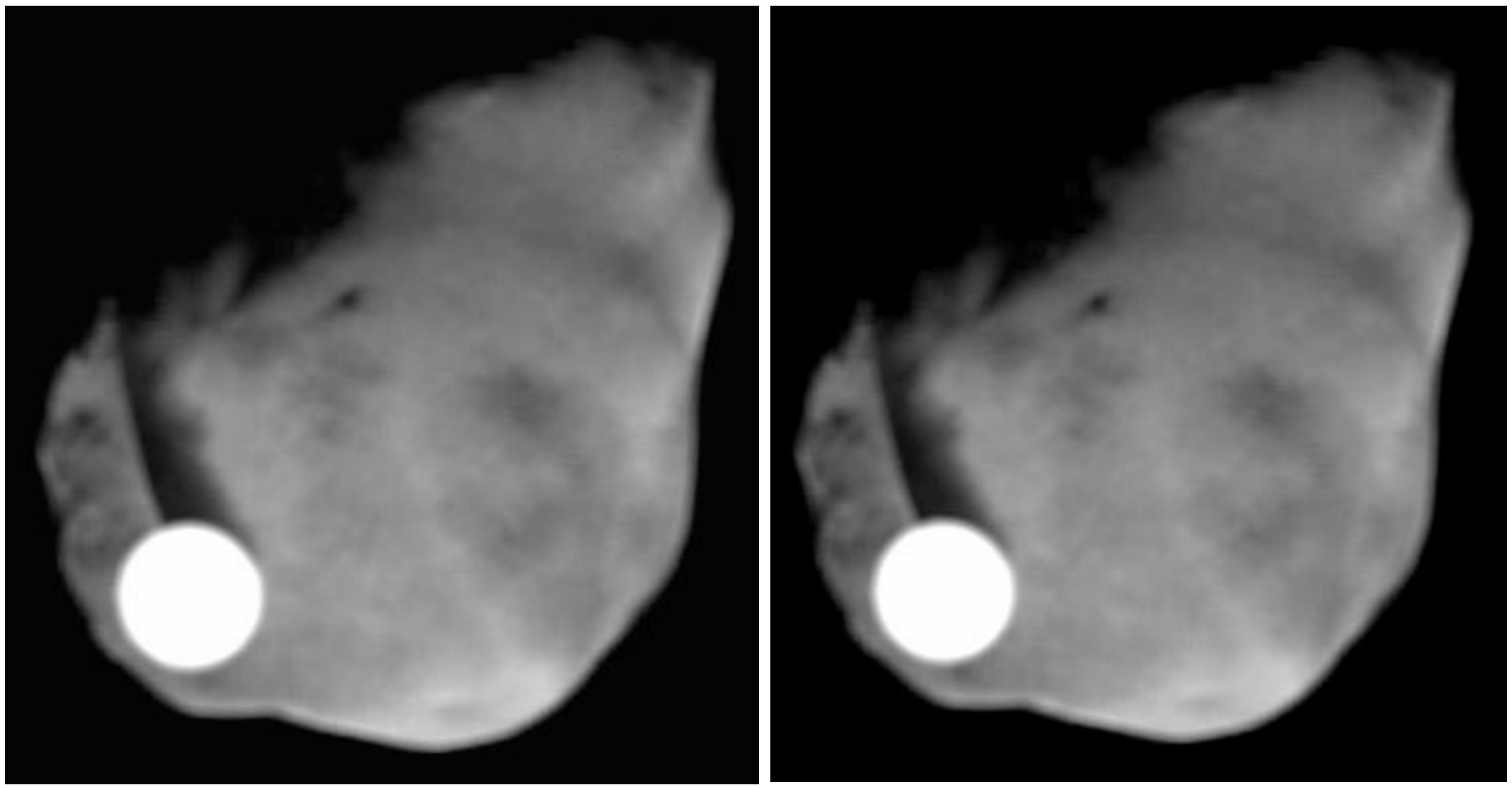

**Figure 3.** Simulated images for the individual particle density of 0.4 g/cm$^3$ (left) and 1.75 g/cm$^3$ (right). In both cases the peak brightness of the ejecta is 9 W/(m$^2$ sr μm) and of the shadow 0.2 W/(m$^2$ sr μm), which reproduces the corresponding peak brightness of the HRI images acquired at 1 second after Deep Impact (panel (a) in Figure 1).